# Simulation of a Non-Newtonian drop impact on a rigid surface: A mess-free approach

Tapan Jana[1*], Amit Shaw[1], L. S. Ramachandra[1]

[1]Department of Civil Engineering, Indian Institute of Technology Kharagpur, India

```
*Email id - tapan.jucivil@gmail.com
```

**Abstract.** The present study explores the impact of a non-Newtonian fluid drop on a rigid surface using a mesh-free approach, Smoothed Particle Hydrodynamics. The complex interaction between viscous and elastic forces during drop impact may be reproduced by integrating the Oldroyd-B model, which describes viscoelastic fluids, into the developed computational framework. A suitable boundary condition is assumed for interaction between fluid and solid surfaces. The ongoing study examines how a droplet deforms, spreads, and recoils after an impact due to its viscoelasticity. The developed computational framework is validated by comparing results from existing literature, providing an understanding of how viscoelasticity influences drop-impact behaviour and opening up new avenues for subsequent research in fluid dynamic problems.

**Keywords:** Drop impact, Non-Newtonian fluid, Oldroyd-B model, Mess-free method, Smoothed Particle Hydrodynamics

## 1  Introduction

The influence of liquid droplets on surfaces is essential in several scientific and commercial applications, such as inkjet printing, spray coating, and microelectronics fabrication. The behaviour of droplets upon contact significantly affects outcomes such as coating effectiveness, liquid absorption, surface retention, and solidification patterns. The quality and accuracy of the procedure are directly influenced by the maximal spread radius, which defines the region encompassed by a droplet. This is a critical factor that affects these phenomena. Although Newtonian fluids have provided a substantial comprehension of droplet impact dynamics, non-Newtonian fluids offer additional complexity due to their unique rheological characteristics, such as shear thinning, shear thickening, and viscoelasticity. These attributes significantly alter the behaviours of spilling, spreading, and absorbing upon impact. Applications in soil erosion management, precision coating, and advanced manufacturing technologies are contingent upon the interaction between the fluid's properties and the surface features.

Large deformations are notoriously difficult to model with FEM because of mesh distortion; the situation becomes worse when the materials in question behave like fluids. The internal limits of FEM still remain, even if adaptive meshing approaches have been created to fix these problems. Recently, Smoothed Particle Hydrodynamics (SPH) has become more well-known as a reliable numerical approach for issues with large deformations and a lot of material flow. For such complicated events, SPH is ideal because of its meshless and adaptable characteristics.

The study of droplet impact dynamics using SPH has advanced significantly, addressing various challenges and unveiling key insights into fluid behavior. Early work by Fang et al. [1] and Jiang et al. [2] focused on modeling viscoelastic and Oldroyd-B fluid impacts, addressing tensile instability through artificial stress, and validating results against experimental and numerical benchmarks. Zhang [3] expanded this research by introducing a boundary-handling technique to simulate surface tension effects in 2D and 3D scenarios. Subsequent studies, such as those by Meng et al. [4] and Ma et al. [5], explored droplet-solid interactions, highlighting the influence of size ratios, impact velocity, and elasticity on spreading and film thickness. Further, Yang et al. [6,7] investigated the effects of impact velocity, thermal conditions, and surface properties on crown formation, spreading, and recoil behavior. More recent works, including Subedi et al. [8], examined the influence of impact frequency and ambient conditions on splashing and lamella dynamics. Collectively, these studies demonstrate SPH's efficacy in capturing complex phenomena such as free surface flows, nonlinear deformations, and thermal effects, paving the way for refined modeling techniques and practical applications. The present work aims to develop a computational framework based on SPH to study the behavior of a non-Newtonian fluid drop impact.

## 2    Mathematical model

SPH, a notable mesh-free methodology developed by Lucy [9] and Gingold [10] for astrodynamical simulations, has become a flexible instrument across several scientific fields. Its application encompasses free surface flows [11-13], impact mechanics [14-15], fracture mechanics [16-17], fluid impact modelling [18], slope-stability analysis [19], and the fracture analysis of concrete dams subjected to dynamic loadings [20]. This highlights SPH's substantial influence and extensive applicability in scientific research. The equations representing the conservation of mass, momentum, and energy, using a Lagrangian framework, are articulated in Einstein's index notation as,

$$\frac{d\rho}{dt} = -\rho \frac{\partial v^\beta}{\partial x^\beta} \qquad (1)$$

$$\frac{dv^\alpha}{dt} = \frac{1}{\rho} \frac{\partial \sigma^{\alpha\beta}}{\partial x^\beta} + g^\alpha \qquad (2)$$

$$\frac{de}{dt} = \frac{\sigma^{\alpha\beta}}{\rho}\frac{\partial v^\beta}{\partial x^\beta} \tag{3}$$

Let the computational domain comprises $N$ number of particles, with their position at $\{x_i\}_{i=1}^N$. Now, at each particle level, i.e., the discrete representations are formulated as

$$\frac{d\rho_i}{dt} = \sum_{j \in N} m_j(v_i^\beta - v_j^\beta) W_{ij,\beta} \tag{4}$$

$$\frac{dv_i^\alpha}{dt} = \sum_{j \in N} m_j \left( \frac{\sigma_i^{\alpha\beta}}{\rho_i^2} + \frac{\sigma_j^{\alpha\beta}}{\rho_j^2} - \Pi_{ij}\delta^{\alpha\beta} \right) W_{ij,\beta} + g^\alpha \tag{5}$$

$$\frac{de_i}{dt} = -\frac{1}{2}\sum_{j \in N} m_j(v_i^\beta - v_j^\beta)\left( \frac{\sigma_i^{\alpha\beta}}{\rho_i^2} + \frac{\sigma_j^{\alpha\beta}}{\rho_j^2} - \Pi_{ij}\delta^{\alpha\beta} \right) W_{ij,\beta} \tag{6}$$

In Equation (5), (6), $\Pi_{ij}$ is artificial viscosity. In the presence of shock, an inaccurate numerical solution can be seen. To stabilize the numerical solution by diffusing the discontinuities created by the shock, artificial viscosity [21] is applied in the present work. It is expressed as

$$\Pi_{ij} = \begin{cases} \dfrac{-\gamma_1 \bar{c}_{ij}\mu_{ij} + \gamma_2 \mu_{ij}^2}{\bar{\rho}_{ij}}, & \text{for } v_{ij}.x_{ij} < 0 \\ 0, & \text{otherwise} \end{cases} \tag{7}$$

Here, $i, j$ are the $i^{th}$ and $j^{th}$ particle respectively. $\mu_{ij} = \frac{h(v_{ij}.x_{ij})}{|r_{ij}|^2 + \epsilon h^2}$, average wave speed, $\bar{c}_{ij} = \frac{c_i + c_j}{2}$ is the average sound speed, $\bar{\rho}_{ij} = \frac{\rho_i + \rho_j}{2}$ is the average density and, $\gamma_1, \gamma_2$ are the artificial viscosity coefficients. The parameter $\epsilon$ is 0.01 ($0 < \epsilon < 1$).

Hydrostatic pressure and deviatoric stress make up the stress tensor. The solvent contribution ($\tau_s$) and the polymeric contribution ($\tau_p$) are the two halves of the deviatoric stress for viscoelastic fluids. Here is one way to describe this relationship:

$$\sigma^{\alpha\beta} = -P\delta^{\alpha\beta} + \tau_s^{\alpha\beta} + \theta.\tau_p^{\alpha\beta} \tag{8}$$

The solvent contribution ($\tau_s$) is represented as

$$\tau_{s,i}^{\alpha\beta} = \eta_s\bigl(k_i^{\alpha\beta} + k_i^{\beta\alpha}\bigr) \tag{9}$$

$$\frac{d\tau_{p,i}^{\alpha\beta}}{dt} = k_i^{\alpha\gamma}\tau_{p,i}^{\gamma\beta} + k_i^{\beta\gamma}\tau_{p,i}^{\gamma\alpha} - \frac{1}{\lambda_1}\tau_{p,i}^{\alpha\beta} + \frac{\eta_p}{\lambda_1}\bigl(k_i^{\alpha\beta} + k_i^{\beta\alpha}\bigr) \tag{10}$$

$$k_i^{\alpha\beta} = \frac{dv_i^\alpha}{dx^\beta} = \sum_{j\in N} m_j/\rho_j (v_i^\beta - v_j^\beta) W_{ij,\beta} \tag{11}$$

Here, $\eta_s$ is solvent viscosity, $\eta_p$ is the polymer contribution to the viscosity, and $\lambda_1$ is the relaxation time of the fluid. The hydrostatic pressure ($P$) is estimated from an equation of state as

$$P = \frac{\rho_0 c_0^2}{\gamma}\left(\left(\frac{\rho}{\rho_0}\right)^\gamma - 1\right) \tag{12}$$

where $c_0$ is sound speed, $\rho_0$ is the reference density, and $\gamma$ is a constant, set to 7. The cubic kernel has been considered for the present work, shown in Fig. 1. The expressions are as follows

$$W(a,h) = \alpha_d \begin{cases} 1 - \frac{3}{2}a^2 + \frac{3}{4}a^3, & \text{if } 0 \leq a \leq 1 \\ \frac{1}{4}(2-a)^3, & \text{if } 1 \leq a \leq 2 \\ 0 & \text{otherwise} \end{cases} \tag{13}$$

where $\alpha_d = \frac{2}{3h}$ in 1D, $\alpha_d = \frac{10}{7\pi h^2}$ in 2D, $\alpha_d = \frac{1}{\pi h^3}$ in 3D, and $a = \frac{|x_i - x_j|}{h}$ represents the normalized position vector linked to a pair of particles.

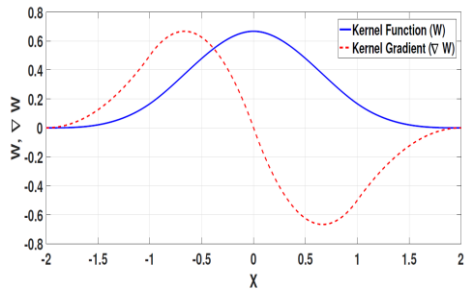

**Fig. 1.** Cubic B-spline Kernel function

Particles should travel in a smooth and well-defined velocity field, keeping their velocities near to the average of their neighbours, to guarantee orderly motion. This was accomplished by adding a correction element to the momentum equation, which Monaghan [22] stated as follows:

$$\frac{dx_i}{dt} = v_i + \varepsilon \sum_{j \in N} \frac{m_j}{\bar{\rho}_{ij}} v_{ij} W_{ij} \qquad (14)$$

Here, $\varepsilon$ is a parameter typically set to 0.5 in most scenarios. In this study, the explicit predictor-corrector time-stepping method is employed to compute the unknown variables based on the discretized governing Equations (4)-(6). The time step is selected to ensure that the information propagation within the domain during each step remains smaller than the inter-particle spacing, adhering to the Courant-Friedrichs-Lewy (CFL) condition. This criterion is expressed as $\Delta t = min_i \left( c_s \frac{h}{c_i + |v_i|} \right)$, where the CFL number ($c_s$) is fixed at 0.3.

## 3  Numerical simulations

A disc-shaped droplet with an initial radius of R = 0.01 m is dropped from a height of H = 0.04 m on a rigid surface with velocity V = 1 m/s (Tomé et al. [23]) (see Fig. 2). Here, we use fixed boundary particles for solid walls and dummy particles in a grid beyond them, inspired by the work of Xu et al. [24]. The pressure of boundary particles is extrapolated from the fluid particles within support domain, and the pressure of the dummy particles is equal to the immediate neighbour boundary particle, ensuring no-slip conditions and enhancing accuracy and stability. The particle spacing is considered at 0.0002 m, which gives 7,957 particles. According to the CFL stability requirements, a time step of 4 x $10^{-6}$ s is used. Other values are: $\eta_s$ = 4.0 Pa-s, $\eta_p$ =3.6 Pa-s, and $\lambda_1$ = 0.02s.

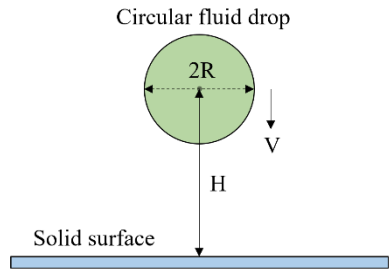

**Fig. 2.** Initial setup of the model

First, the simulation is performed using a conventional SPH approach, and the result is shown in Fig. 3(a). In this result, numerical instability is seen, i.e., tensile instability. It occurs due to small wavelength zero energy modes, which in turn pollutes the whole simulation. In SPH, particles interact through kernel function. When two particles are in tension, due to characteristics of the kernel function (cubic), with the in-creased particle distance, the force between them decreases. This creates negative stiffness, which in turn causes numerical instability. In the next stage, the simulation is performed using an adaptive kernel-based approach. The result, shown in Fig. 3(b), indicates that no instability is present. The issue is mitigated successfully.

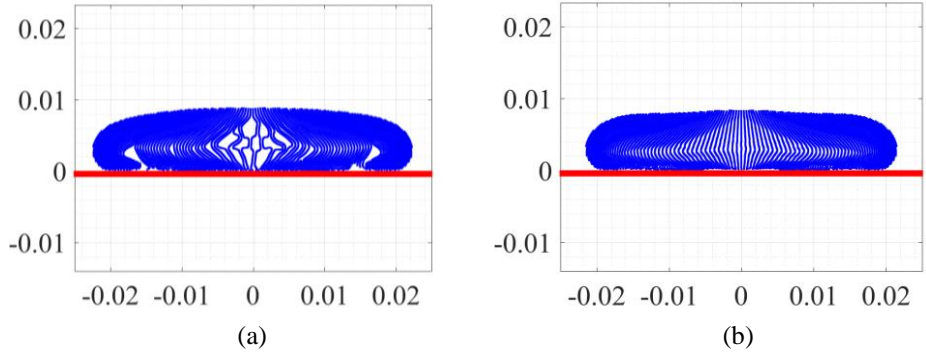

**Fig. 3.** Simulation using (a) conventional SPH and (b) adaptive kernel approach at T = 2.3.

Fig. 4 shows the horizontal velocity/particle positions contour for the drop at non-dimensional times, T = tV/2R. The plots show that after hitting the rigid surface, the drop starts spreading as it flows fast, but later spreading slows down due to the viscosity effect. Over time, due to the elasticity effect, the drop exhibits retraction behaviour characterized by a positive vertical velocity. Later, it loses elasticity and starts spreading over the surface again. The results also indicate that particles on the left and right sides of the drop move in opposite directions with same velocity.

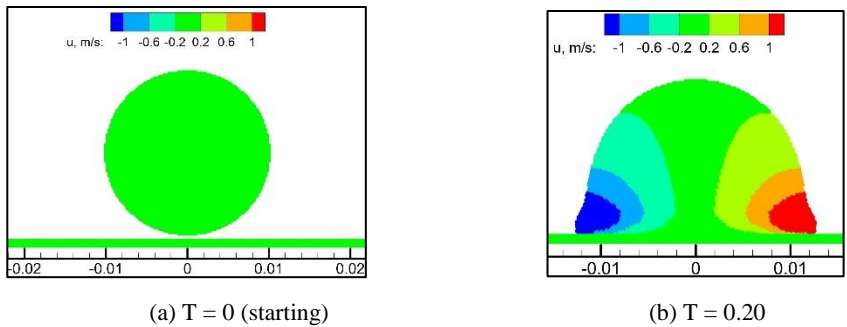

(a) T = 0 (starting)          (b) T = 0.20

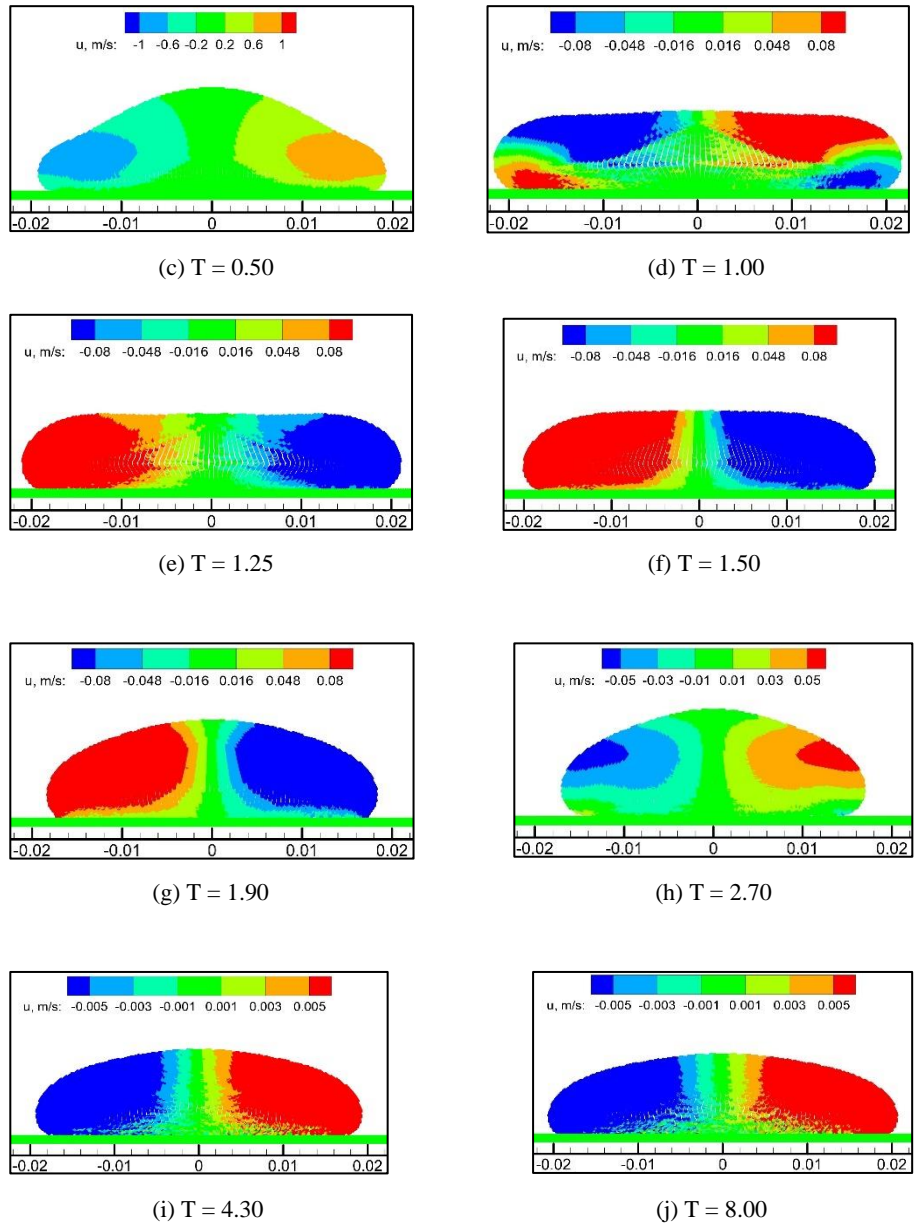

**Fig. 4.** Time evolution of the drop: the horizontal velocity/particle positions contour

Now, the time history of the fluid drop's width, compared with the FDM results (Tomé et al. [23]) in Fig. 5, shows a good agreement with the existing literature.

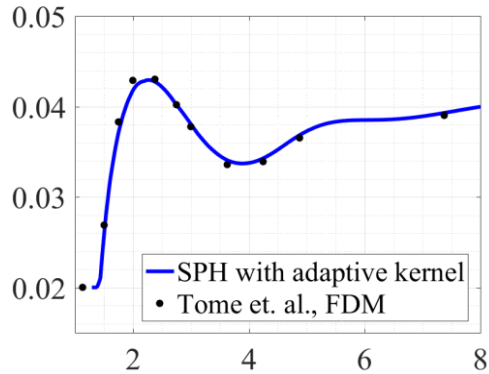

**Fig. 5.** Comparison of drop width time history between the adaptive kernel-based simulated result with the FDM results from the work of Tomé et al. [23]

## 4 Conclusion

Droplet impact is a complicated phenomenon and a major real-world challenge. The present work uses an SPH-based computational approach to simulate it efficiently. This study demonstrates how the SPH approach can capture important aspects of droplet impact dynamics and improve our knowledge of the phenomena. Additionally, the tensile instability issue is also addressed. The successful validation against the existing literature further strengthens the robustness and reliability of the computational framework.